\documentclass[twocolumn]{aastex61}

\pdfoutput=1

\usepackage{enumerate}
\usepackage{natbib,ifthen}
\usepackage{graphicx}
\usepackage{fancyhdr}
\usepackage{amssymb,amsmath}
\usepackage{multirow,bigdelim}
\usepackage{chngcntr}

\shorttitle{Complex Organic Molecules in the LMC}
\shortauthors{Sewi{\l}o et al.}

\begin{document}


\title{The Detection of Hot Cores and Complex Organic Molecules in the Large Magellanic Cloud}


\correspondingauthor{Marta Sewi{\l}o}
\email{marta.m.sewilo@nasa.gov}

\author{Marta Sewi{\l}o}
\altaffiliation{NASA Postdoctoral Program Fellow}
\affiliation{NASA Goddard Space Flight Center, 8800 Greenbelt Rd, Greenbelt, MD 20771, USA}

\author{Remy Indebetouw}
\affiliation{Department of Astronomy, University of Virginia, PO Box 400325, Charlottesville, VA 22904, USA}
\affiliation{National Radio Astronomy Observatory, 520 Edgemont Rd, Charlottesville, VA 22903, USA}

\author{Steven B. Charnley}
\affiliation{NASA Goddard Space Flight Center, 8800 Greenbelt Rd, Greenbelt, MD 20771, USA}

\author{Sarolta Zahorecz}
\affiliation{Department of Physical Science, Graduate School of Science, Osaka Prefecture University, 1-1 Gakuen-cho, Naka-ku, Sakai, Osaka 599-8531, Japan}
\affiliation{Chile Observatory, National Astronomical Observatory of Japan, National Institutes of Natural Science, 2-21-1 Osawa, Mitaka, Tokyo 181-8588, Japan}

\author{Joana M. Oliveira}
\affiliation{Lennard-Jones Laboratories, Keele University, ST5 5BG, UK}

\author{Jacco Th. van Loon}
\affiliation{Lennard-Jones Laboratories, Keele University, Staffordshire ST5 5BG, UK}

\author{Jacob L. Ward}
\affiliation{Astronomisches Rechen-Institut, Zentrum f{\"u}r Astronomie der Universit{\"a}t Heidelberg, M{\"o}nchhofstr. 12-14, 69120 Heidelberg Germany}

\author{C.-H. Rosie Chen}
\affiliation{Max-Planck-Institut f{\"u}r Radioastronomie, Auf dem H{\"u}gel 69 D-53121 Bonn, Germany}

\author{Jennifer Wiseman}
\affiliation{NASA Goddard Space Flight Center, 8800 Greenbelt Rd, Greenbelt, MD 20771, USA}

\author{Yasuo Fukui}
\affiliation{School of Science, Nagoya University, Furo-cho, Chikusa-ku, Nagoya 464-8602, Japan}

\author{Akiko Kawamura}
\affiliation{National Astronomical Observatory of Japan, 2-21-1 Osawa, Mitaka, Tokyo 181-8588, Japan}

\author{Margaret Meixner}
\affiliation{Space Telescope Science Institute, 3700 San Martin Drive, Baltimore, MD 21218, USA}

\author{Toshikazu Onishi}
\affiliation{Department of Physical Science, Graduate School of Science, Osaka Prefecture University, 1-1 Gakuen-cho, Naka-ku, Sakai, Osaka 599-8531, Japan}

\author{Peter Schilke}
\affiliation{I. Physikalisches Institut der Universit{\"a}t zu K{\"o}ln, Z{\"u}lpicher Str. 77, 50937, K{\"o}ln, Germany}


\begin{abstract}
We report the first extragalactic detection of the complex organic molecules (COMs) dimethyl ether (CH$_3$OCH$_3$) and methyl formate (CH$_3$OCHO) with the Atacama Large Millimeter/submillimeter Array (ALMA). These COMs together with their parent species methanol (CH$_3$OH), were detected toward two 1.3 mm continuum sources in the N\,113 star-forming region in the low-metallicity Large Magellanic Cloud (LMC).  Rotational temperatures ($T_{\rm rot}\sim130$~K) and total column densities ($N_{\rm rot}\sim10^{16}$~cm$^{-2}$) have been calculated for each source based on multiple transitions of CH$_3$OH.  We present the ALMA molecular emission maps for COMs and measured abundances for all detected species.  The physical and chemical properties of two sources with COMs detection, and the association with H$_2$O and OH maser emission indicate that they are hot cores.  The fractional abundances of COMs scaled by a factor of 2.5 to account for the lower metallicity in the LMC are comparable to those found at the lower end of the range in Galactic hot cores.  Our results have important implications for studies of organic chemistry at higher redshift.
\end{abstract}



\section{Introduction}

Complex organic  molecules (COMs, $\ge$6 atoms including carbon; e.g., CH$_3$OH, CH$_3$OCH$_3$, CH$_3$OCHO) are widespread in the Milky Way galaxy where they  have primarily been found in the environments of young protostars -- hot cores and hot corinos  (e.g.,  \citealt{herbst2009}). Hot cores are compact ($D\lesssim0.1$~pc), hot ($T_{\rm kin}>100$~K), dense ($n_{\rm H}>10^{6-7}~$cm$^{-3}$) sources  (e.g., \citealt{kurtz2000}) where ice mantles have recently been removed from dust grains, either by thermal  evaporation or sputtering in shock waves.  Interstellar COMs may be  a chemical link  to the  prebiotic molecules that were involved in the processes leading to the origin of life \citep{ehrenfreund2000}. 

The observed hot core molecules may have their origin in the chemistry of  the cold prestellar phase, either through formation in  gas phase reactions followed by freeze-out on the dust, or as products of  reactions on the icy mantles of the dust grains \citep{brown1988}.  Grain-surface production could involve  atom addition reactions on cold dust (e.g., \citealt{charnley2008}) or radical reactions on warm dust during the thermal warm-up of the hot core \citep{garrod2006}. Alternatively, COMs and other molecules may be formed in subsequent gas-phase reactions in the hot gas (e.g., \citealt{charnley1992}). 

In galaxies with sub-solar metallicities, questions remain as to the formation efficiency of COMs  (e.g., \citealt{acharyya2015};  \citealt{shimonishi2016a}).  Apart from the lower elemental abundances of gaseous  C, O, and N atoms, low metallicity leads to less shielding, greater penetration of UV photons into dense gas, and consequently warmer dust grains (e.g., \citealt{vanloon2010};  \citealt{oliveira2011}). All of these factors inhibit the formation and survival of COMs. 

 The Large Magellanic Cloud (LMC) is a nearby dwarf galaxy ($50.0\pm1.1$~kpc; \citealt{pietrzynski2013})  that presents the opportunity to study COM chemistry in a low-metallicity environment ($Z_{\rm LMC}\sim0.3-0.5\,Z_{\odot}$; \citealt{westerlund1997}) with higher UV fluxes and lower cosmic-ray ionization rates than the Milky Way (e.g., \citealt{abdo2010}).  \citet{shimonishi2016a} have claimed a hot core detection toward the LMC Young Stellar Object (YSO) ST11 based on the derived physical conditions and the presence of simple molecules connected to hot gas chemistry (e.g., SO$_2$ and its isotopologues). Neither methanol or  COMs were detected  \citep{shimonishi2016b}. 
 
 \citet{nishimura2016a} observed  seven molecular clouds in the LMC at low spatial resolution ($\sim$8~pc) and also failed to detect methanol or any COMs.  Several transitions of methanol have previously been reported in two star-forming regions in the LMC (N\,159W and N\,113; \citealt{heikkila1999}; \citealt{wang2009}), both included in the \citet{nishimura2016a} sample, and a few CH$_3$OH masers are reported in literature (e.g., \citealt{sinclair1992}; \citealt{green2008}). The existing studies indicate that the abundance of CH$_3$OH in the LMC is very low, with potentially important general implications for COM formation in low-metallicity galaxies and at high redshift. 

The sources observed by \citet{nishimura2016a} contain large abundances of CCH, a good photodissociation region (PDR) tracer, with the highest CCH column density found in N\,113.  This suggests that the formation of COMs in the LMC and other low-metallicity galaxies (e.g.,  IC10, \citealt{nishimura2016b}),  could be inhibited by the presence of extensive PDRs  (\citealt{hollenbach1999}). Apart from destruction by photolysis, warm dust ($\gtrsim$20~K) leads to inefficient H atom sticking and CO hydrogenation to methanol on grain surfaces (e.g., \citealt{watanabe2008}). 

In this Letter, we report the first extragalactic detection of the COMs dimethyl ether (CH$_3$OCH$_3$)  and  methyl formate (CH$_3$OCHO) in the N\,113 region of the LMC. We present the ALMA molecular emission maps and measured abundances of the COMs.  The relationship to Galactic hot cores and the implications for extragalactic organic chemistry are briefly discussed.                

\begin{figure*}
\centering
\includegraphics[width=\textwidth]{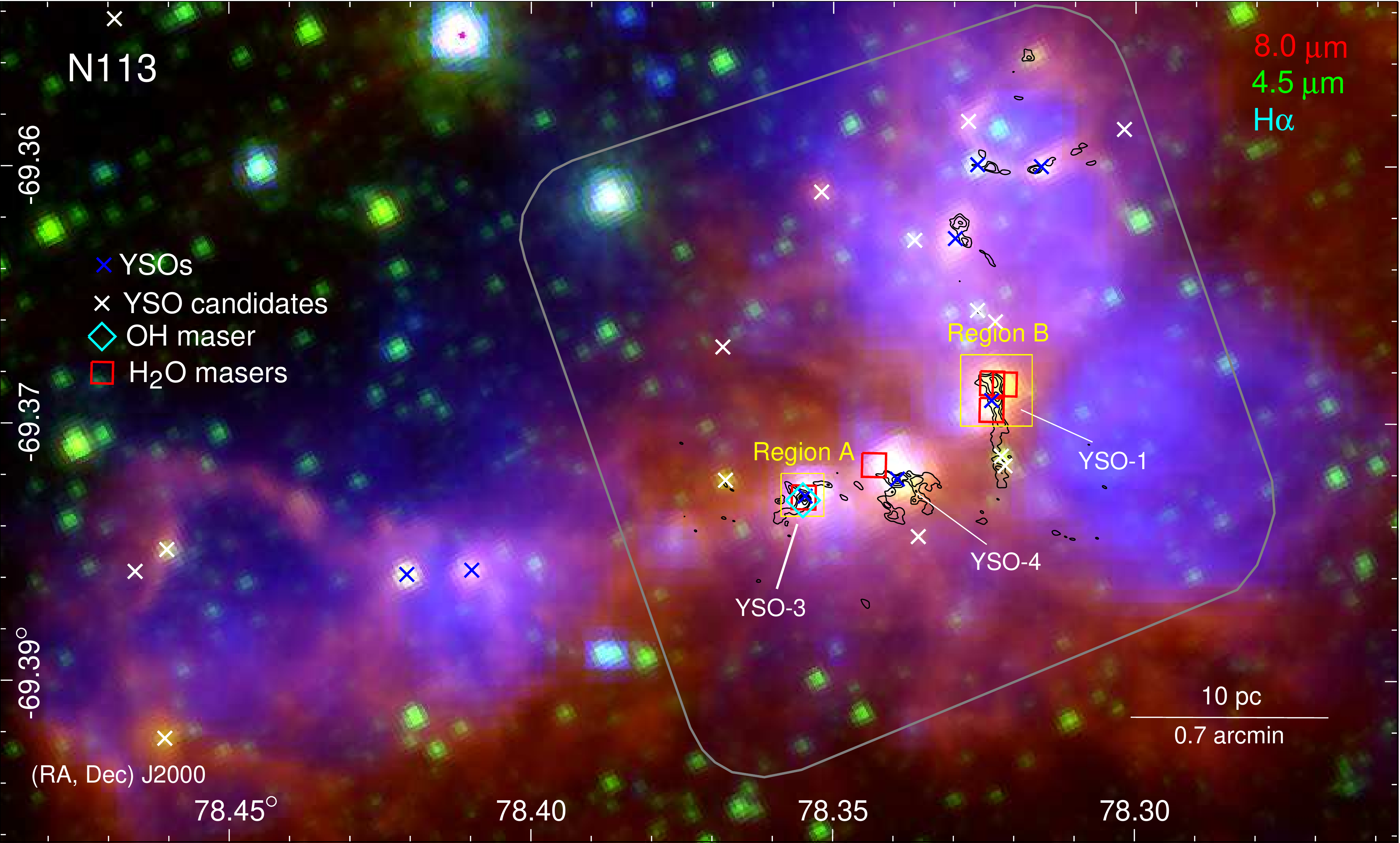}
\caption{The three-color composite image of N\,113 combining the {\it Spitzer}/SAGE IRAC 8.0 $\mu$m ({\it red}) and 4.5 $\mu$m ({\it green}; \citealt{meixner2006}), and  the MCELS H$\alpha$ ({\it blue}; \citealt{smith1998}) images.  The grey rounded rectangle shows an approximate ALMA Band 6 coverage. The positions of the YSOs, YSO candidates, and masers are marked as indicated in the legend.  Black contours correspond to the ALMA $^{13}$CO (2--1) integrated intensity with contour levels of (10, 20, 40, 60, 80)\% of the peak of  3.17 Jy beam$^{-1}$ km s$^{-1}$.  Regions A and B where COMs are detected are indicated with yellow boxes. Source YSO--2 is outside the field of view. \label{f:n113} }
\end{figure*}


\section{The N\,113 Star-Forming Region}

LHA\,120--N\,113 (N\,113; \citealt{henize1956}) is one of the most prominent star formation regions in the LMC.  It contains one of the most massive ($\sim$10$^{5}$ M$_{\odot}$) and richest giant molecular clouds (GMCs) in the LMC.  Its peak CO (1--0) brightness temperature of $\sim$8.1 K is the highest in the MAGMA survey with the Mopra telescope (half-power beam width, HPBW$\sim$45$''$; \citealt{wong2011}). 

The dense molecular gas in the N\,113 GMC is clumpy with substructures that are directly revealed by the high volume density tracers HCO$^{+}$ and HCN (e.g., \citealt{seale2012}). N\,113 hosts the largest number of H$_{2}$O and OH masers and  the brightest H$_{2}$O maser in the entire LMC (e.g., \citealt{whiteoak1986}; \citealt{green2008}; \citealt{ellingsen2010}), all located within 25$''$ ($\sim$6 pc) from the GMC's CO (1--0) peak. 

Current star formation activity in N\,113 appears concentrated in the central part of the GMC as indicated by point-like mid-infrared emission, maser sources, and compact H\,{\sc ii} regions, superposed on an extended emission component and aligned in a northwest-southeast direction (see Fig.~\ref{f:n113}).  The gas and dust in this central region are compressed by a complex structure of ionized gas bubbles (prominent in the H$\alpha$ images) created by massive stars in several young clusters \citep{oliveira2006}. Based on the {\it Spitzer} Space Telescope and the {\it Herschel} Space Observatory data, confirmed and candidate Stage 0--II YSOs down to $\sim$3 M$_{\odot}$ were identified  (e.g., \citealt{sewilo2010}; \citealt{carlson2012}).

At {\it Spitzer} and {\it Herschel} wavelengths (3.6--500 $\mu$m with 2$''$--38$''$ resolution), N\,113 is dominated by emission from 3 massive (30--40 M$_{\odot}$; \citealt{ward2016}) YSOs associated with high-density molecular clumps, radio continuum emission, and masers: YSO--1 (051317.69--692225.0; \citealt{gruendl2009}), YSO--3 (051325.09--692245.1), and YSO--4 (051321.43--692241.5); see Fig.~\ref{f:n113}.  Our {\it Herschel} spectroscopic data (e.g., [O\,{\sc i}], [O\,{\sc iii}], [C\,{\sc ii}]) reveal distinct physical conditions around each of these close-by sources (J. M. Oliveira et al., in prep.).  {\it K}-band observations obtained with VLT/SINFONI with 0$\rlap.{''}$1 resolution and a field of view $\sim$3$''$ resolved YSO--3 and YSO--4 into a cluster of multiple components and revealed diverse characteristics of resolved sources \citep{ward2016}. N\,113 is a complex environment where we find clumps ($\sim$1~pc) and cores ($\sim$0.1--0.2~pc) in a range of environments (e.g., strongly influenced by the stellar winds in the west and more quiescent in the east) and at a range of evolutionary stages. 

\begin{figure*}
\centering
\hspace*{0.6cm}
\includegraphics[width=0.3\textwidth]{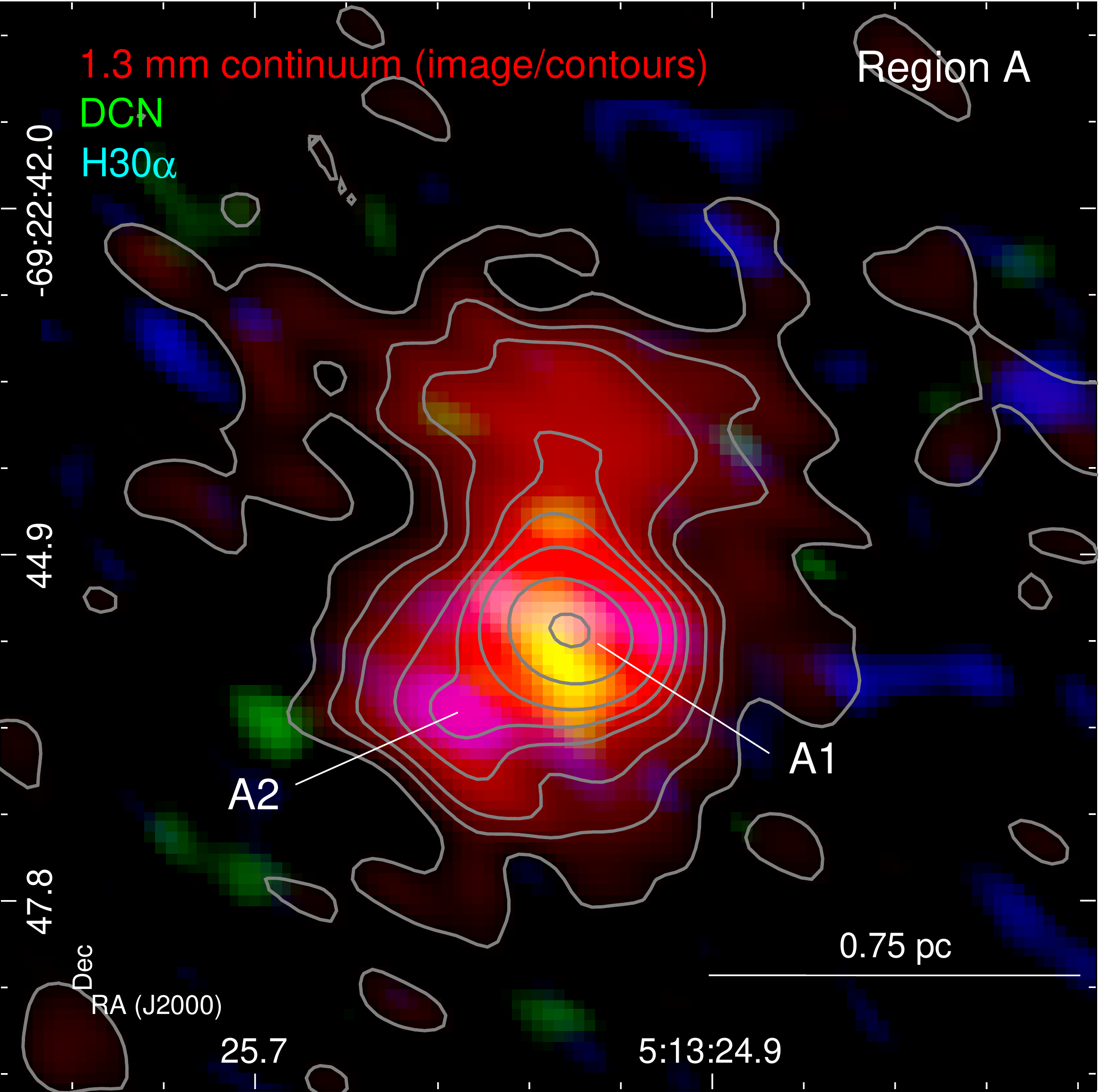} \hspace*{1.5cm}
\includegraphics[width=0.3\textwidth]{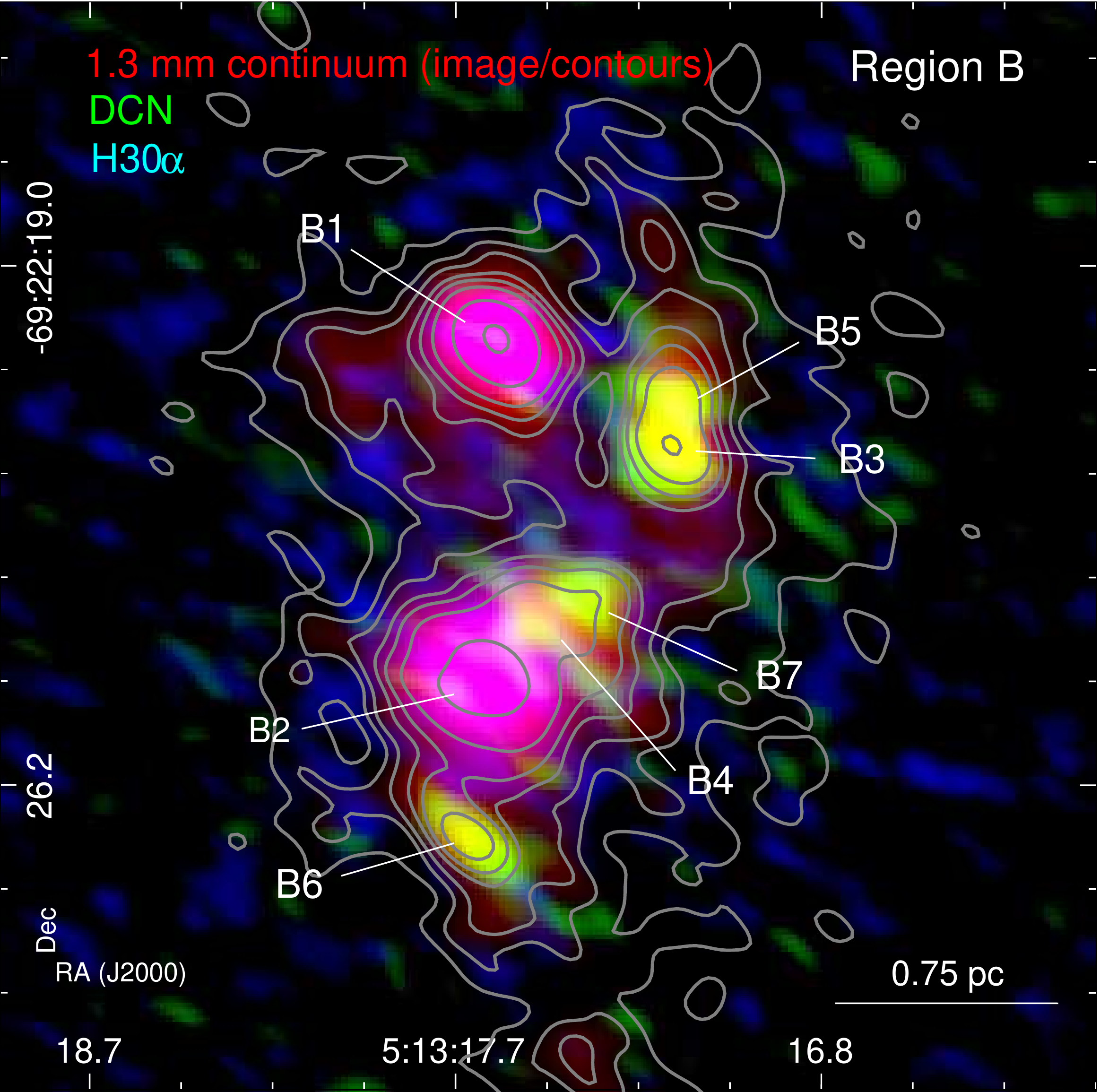} 
\includegraphics[width=0.4\textwidth]{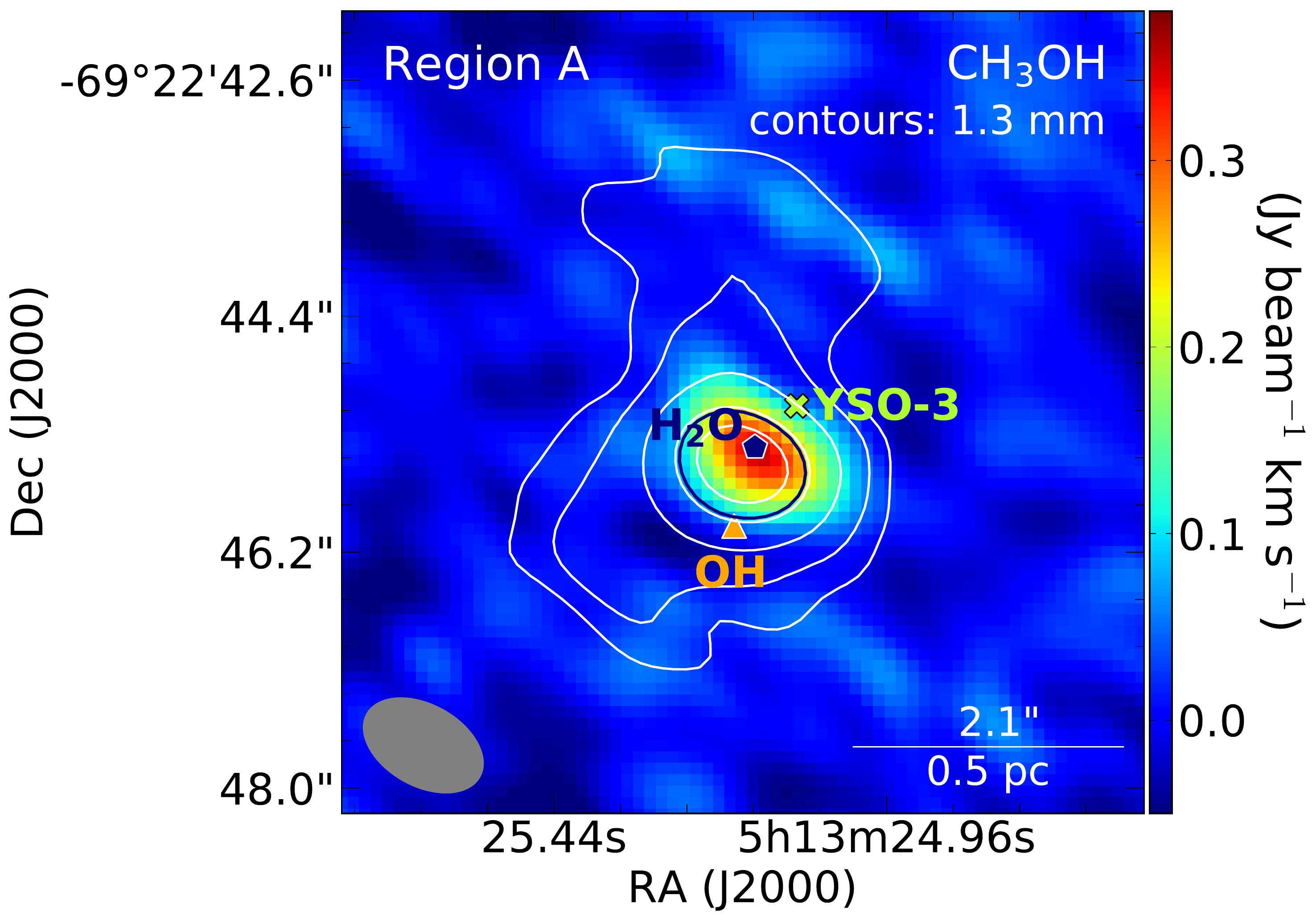}
\includegraphics[width=0.4\textwidth]{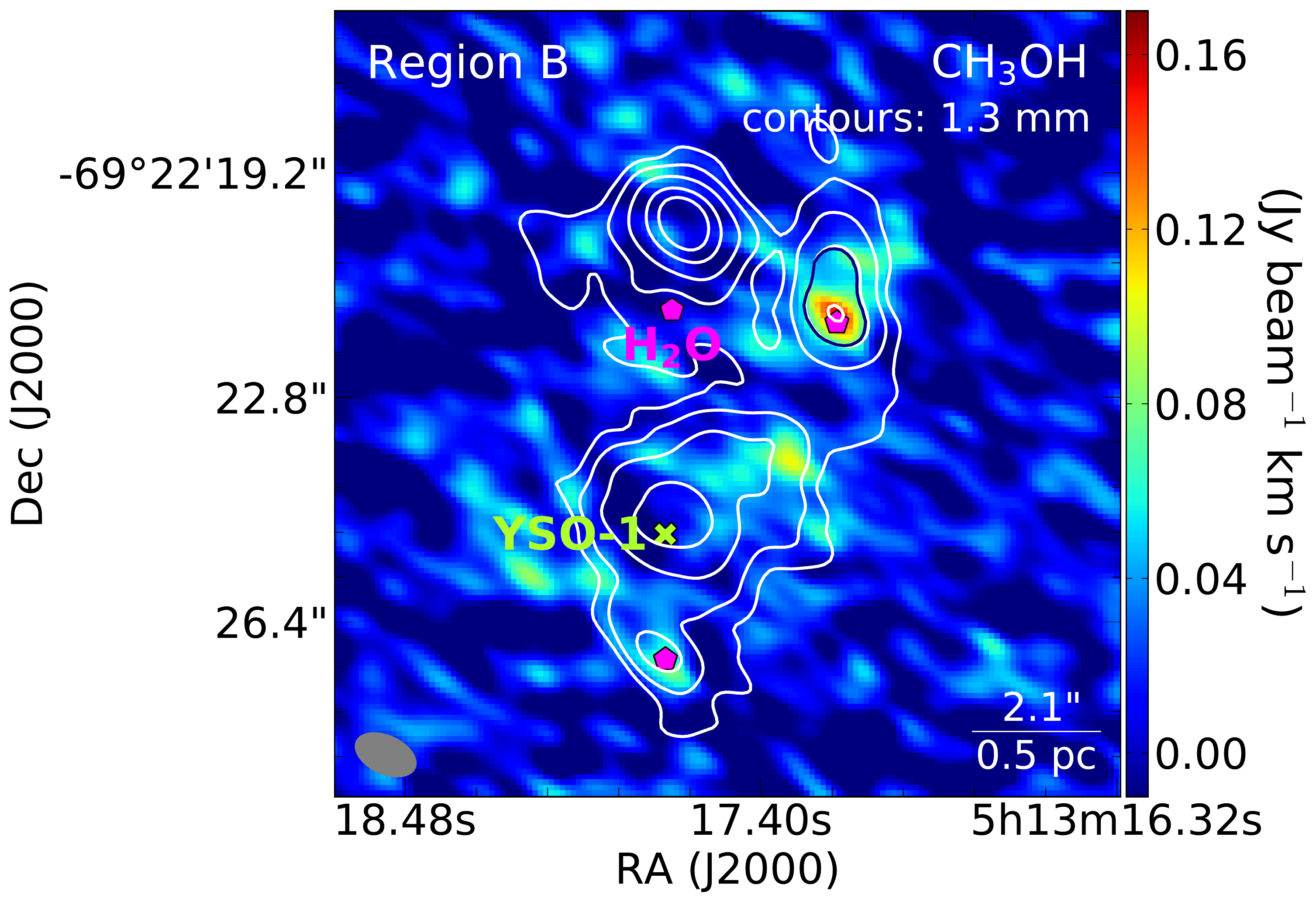} 
\includegraphics[width=0.4\textwidth]{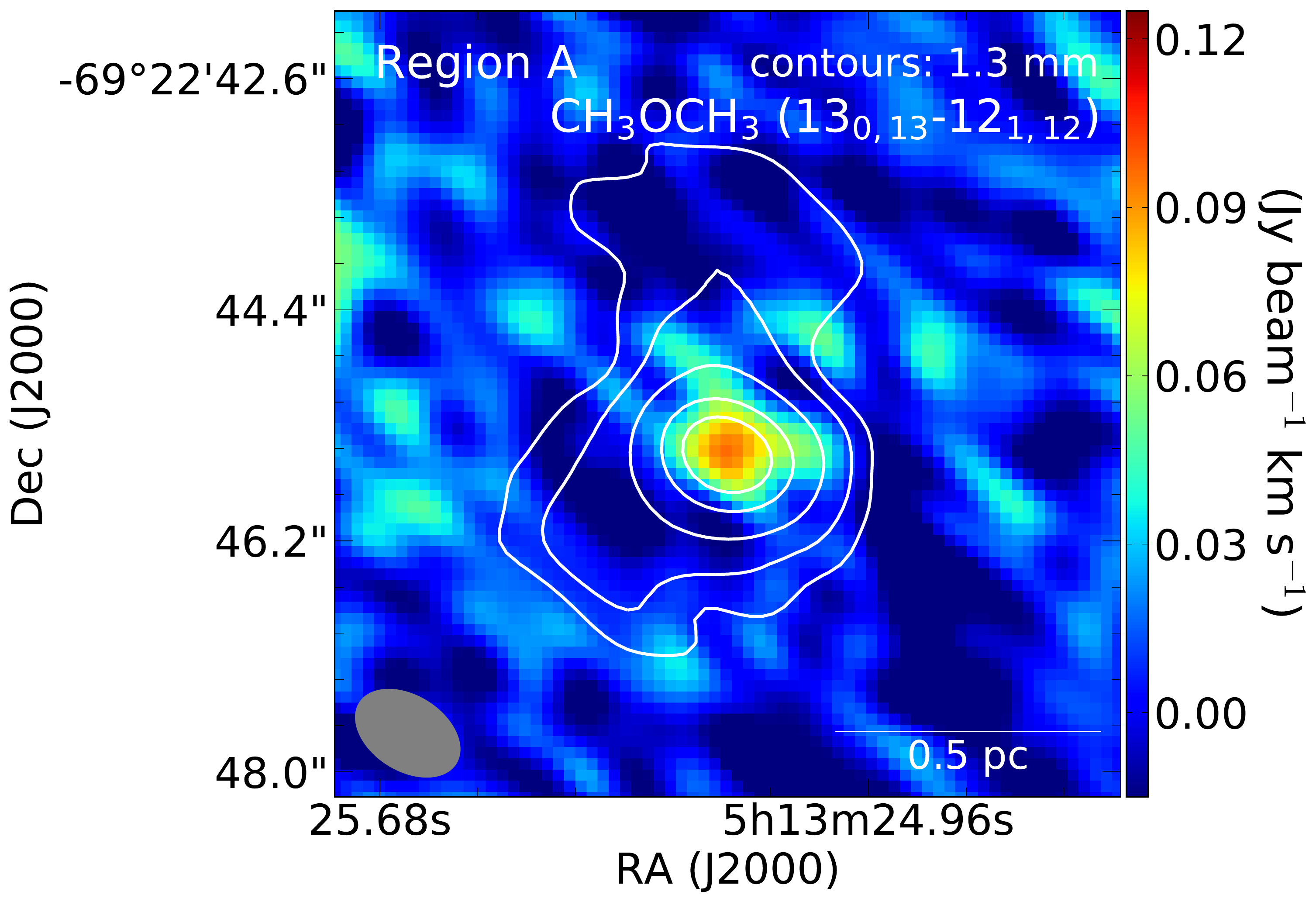}
\includegraphics[width=0.4\textwidth]{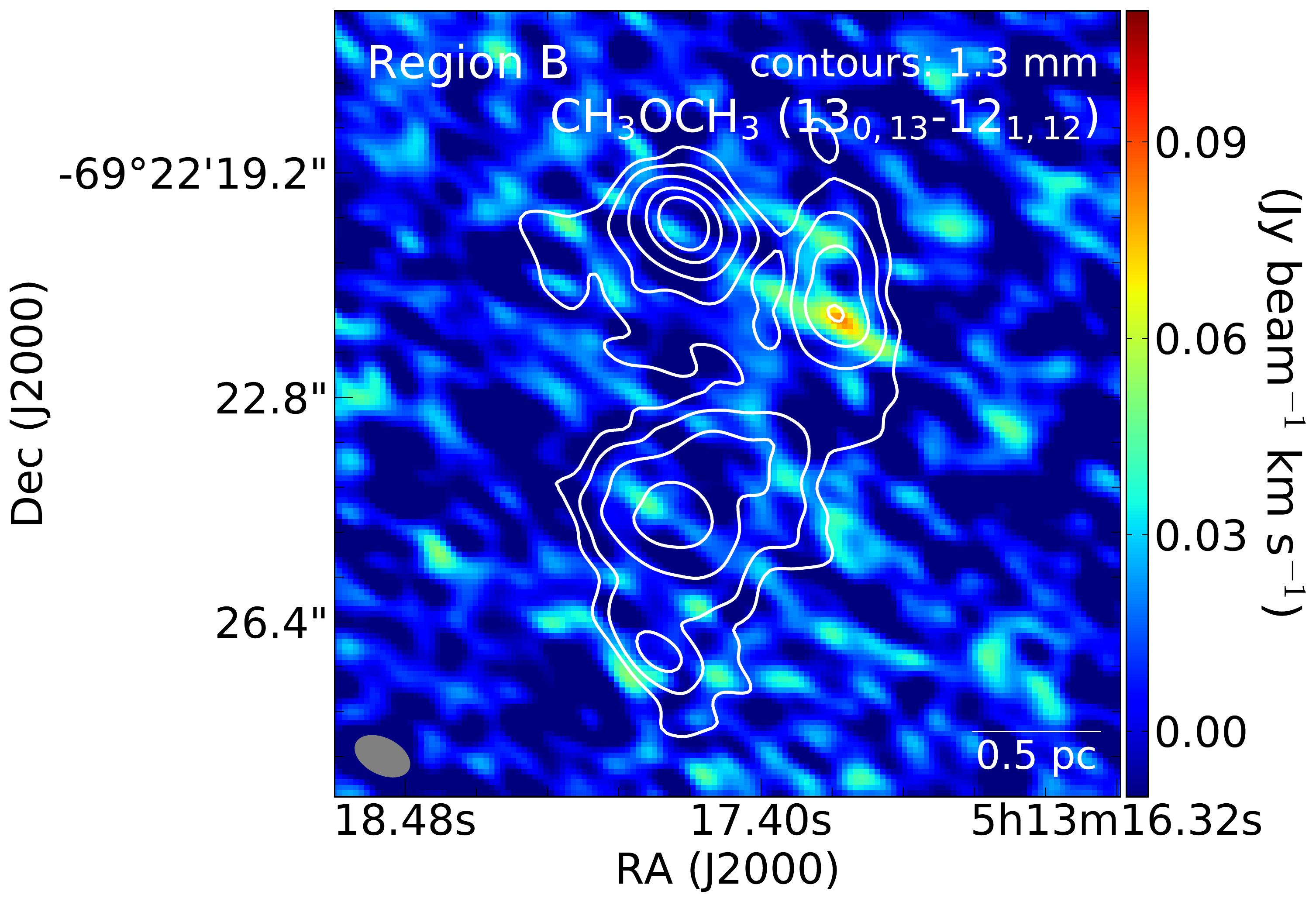}
\includegraphics[width=0.4\textwidth]{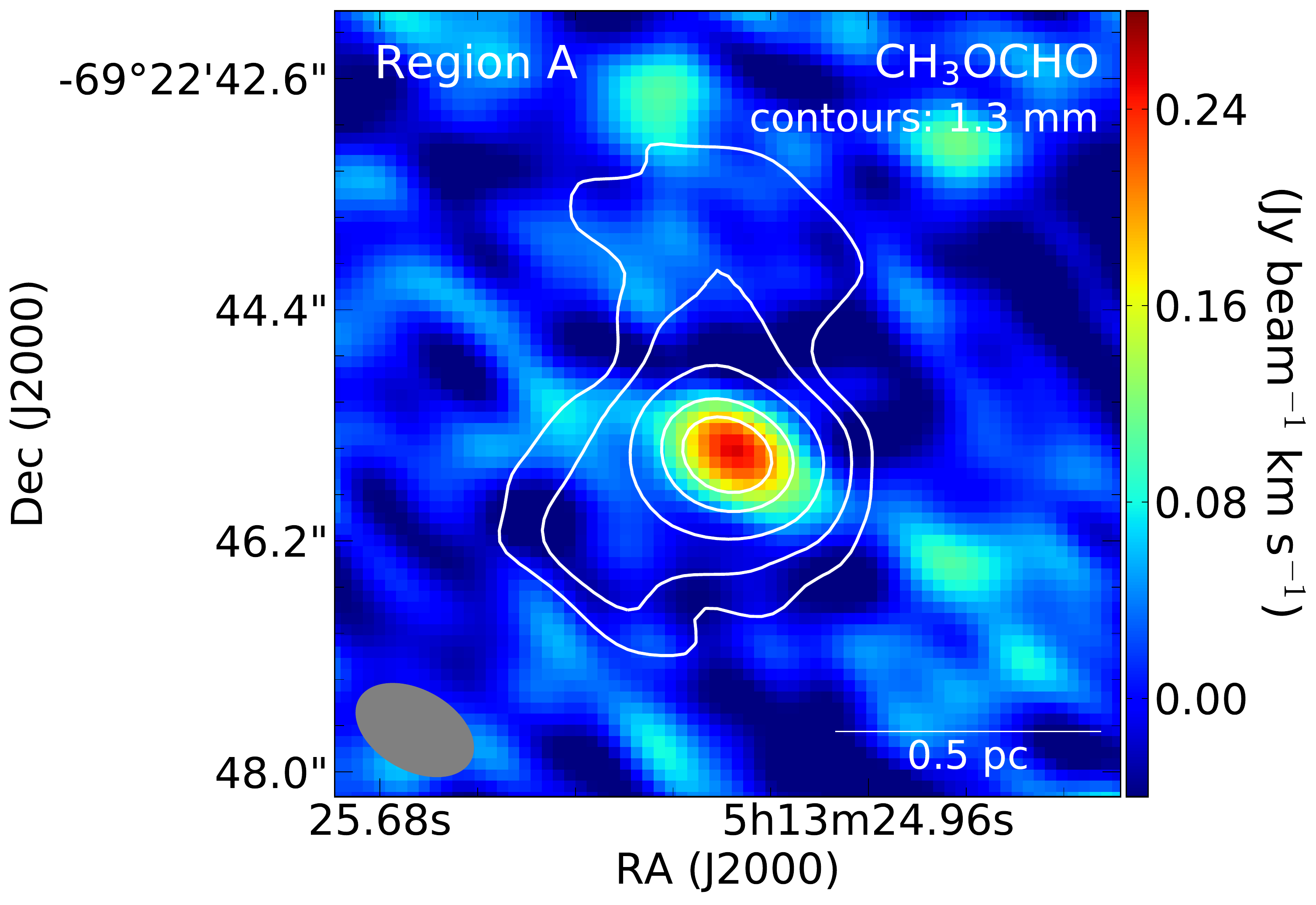}
\includegraphics[width=0.4\textwidth]{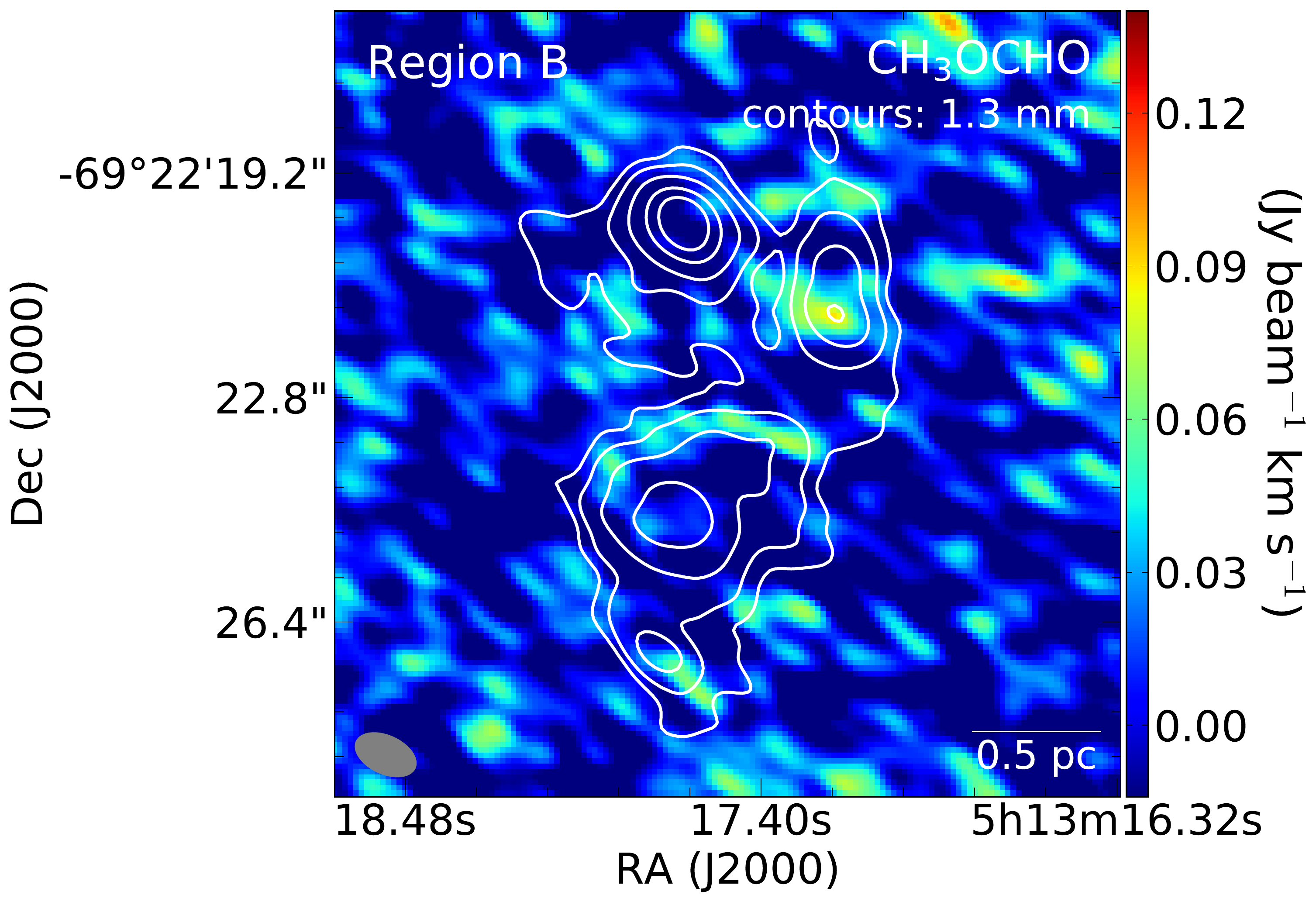}
\caption{{\it Top panel:}  The three-color mosaics combining the 1.3 mm continuum ({\it red}), DCN ({\it green}), and H30$\alpha$ ({\it blue}) images for Region A ({\it left}) and B ({\it right}).  The 1.3 mm continuum contour levels are $(3,6,9,15,20,30,60,120)\times0.1$ mJy~beam$^{-1}$ (1$\sigma$).  {\it Second row:} The CH$_3$OH integrated intensity image for Region A and B with selected 1.3 mm continuum contours overlaid for reference.  The  CH$_3$OH images were made using the channels corresponding to all CH$_3$OH transitions in the 216.9 GHz spectral window.  The positions of  the H$_2$O ({\it pentagons}) and OH ({\it triangle}) masers and  {\it Spitzer} YSOs (`$\times$') are indicated.  The areas enclosed by navy contours were used to extract the spectra shown in Fig.~\ref{f:spec}. The CH$_3$OCH$_3$  (13$_{0,13}$--12$_{1,12}$) and CH$_3$OCHO (integrated over all transitions) images for Regions A and B are shown in the {\it third} and {\it fourth rows}. \label{f:molimages}}
\end{figure*}


\section{ALMA Observations of N\,113}

N\,113 was observed with ALMA in Band 6 as part of project 2015.1.01388.S.  The spectral setup observed $^{12}$CO\,(2--1), $^{13}$CO\,(2--1), and C$^{18}$O\,(2--1) in narrow windows, and additional 2 GHz windows centered on 231.7 and 216.9~GHz.  

The data were obtained with the 12m array on March 10 and June 16, 2016, for a total of 13.1 min per mosaic pointing over most of the map, on baselines from 15 to 704 m.  It was also observed 19 times with the 7m Atacama Compact Array (ACA) array between November 2 and December 17, 2015, for a total of 247 min per mosaic pointing.  The data were calibrated with versions 4.7 through 5.0 of the ALMA pipeline in CASA (Common Astronomy Software Applications; \citealt{mcmullin2007}).  Amplitude was calibrated using Uranus, J0538--4405, J0519--4546, and Callisto (different calibrators on different dates). The bandpass was calibrated using J0538--4405, J0522--3627, J0854$+$2006, J0006--0623, and time-varying gain using J0529--7245 (12m array) and J0440--6952 (7m array).

Sensitivity of 0.1~mJy per $0\rlap.^{''}$87$\times$0$\rlap.^{''}$54 ($\sim$0.21$\times$0.13~pc) beam was achieved in the continuum.  Continuum was subtracted in the {\sc uv} domain from each line spectral window, and the 7m and 12m data were simultaneously imaged and deconvolved interactively.
Sensitivity of 3.8~mJy per 0$\rlap.^{''}$98$\times$0$\rlap.^{''}$62 ($\sim$0.24$\times$0.15~pc) beam was achieved in the 216.9$\;$GHz cube with 1.09$\;$MHz (1.5$\;$km$\;$s$^{-1}$) channels, and 2.9~mJy per 0$\rlap.^{''}$88$\times$0$\rlap.^{''}$58  ($\sim$0.21$\times$0.14~pc) beam in 2.0$\;$MHz (2.6$\;$km$\;$s$^{-1}$) channels at 231.7$\;$GHz.

The distribution of our project's main targeted bright lines over the entire N\,113 region will be discussed in a future publication, as the scope of this Letter is the complex organic molecules located in two small regions in the center of N\,113 (Regions A and B in Fig.~\ref{f:n113}). 


\section{Results}


\subsection{1.3 mm Continuum Emission}
\label{s:continuum}

We have identified multiple 1.3 mm continuum sources in both Region A and B (see Fig.~\ref{f:molimages}). We have assigned identification numbers to all the continuum peaks that are associated with the molecular or H30$\alpha$ line emission in the order from the brightest to faintest. There is a number of the continuum peaks that are less significant and/or not associated with the molecular or H30$\alpha$ line emission that will be discussed in a follow-up paper.  COMs have been detected toward sources A1 and B3. Both sources are associated with H$_2$O masers, and A1 also with an OH maser (within the positional uncertainties).  Only one SINFONI {\it K}-band source (N113--YSO01) has been detected at 1.3 mm (B2).

There is some evidence for a small difference in the evolutionary stage between A1 and B3. B3 is brighter than A1 in DCN, which indicates that it is younger.  The DCN/HCN ratio will drop rapidly with time in hot gas, irrespective of the origin of DCN in the gas phase or on ice. The abundance of all deuterated species decreases with time \citep{fontani2011}; in hot cores, they are fossils of the earlier, colder evolutionary phases.   The more mature nature of A1 is supported by an association with the OH maser.   

We measured the continuum surface brightness for A1 and B3 in the continuum image centered at 224.6 GHz by fitting the sources with Gaussians, as well as measuring the value of the brightest pixel.  The weighted average of the fitted and manually measured peaks ($0.58\pm0.09$~K and $0.29\pm0.02$~K for A1 and B3, respectively) was used to calculate the dust column density.  The dust optical depth was calculated by assuming that the dust temperature equals the fitted CH$_3$OH kinetic temperature (see Section~\ref{s:specline}). The largest source of uncertainty is the choice of dust emissivity used to calculate column density from optical depth.  We use values calculated specifically for the LMC by \citet{galliano2011}, albeit on somewhat larger scales ($\ge$10~pc): $N({\rm H}_2)/\tau_{\rm dust}$ = $1.8\times10^{26}$~cm$^{-2}$.  If dust has coagulated on the small scales probed by ALMA, the dust emissivity could be higher, hydrogen column densities lower, and organic molecule abundances higher than we quote here. We determined $N({\rm H}_2)$  of  $(8.0\pm1.2)\times10^{23}$~cm$^{-2}$  and $(7.0\pm0.9)\times10^{23}$~cm$^{-2}$ for A1 and B3, respectively.  The source diameters are about the beam size, thus under the assumption of spherical symmetry, these column densities correspond to the number densities of $\sim$1.6$\times$10$^6$~cm$^{-3}$ and $\sim$1.4$\times$10$^6$~cm$^{-3}$, respectively.

\begin{figure*}
\centering
\includegraphics[width=\textwidth]{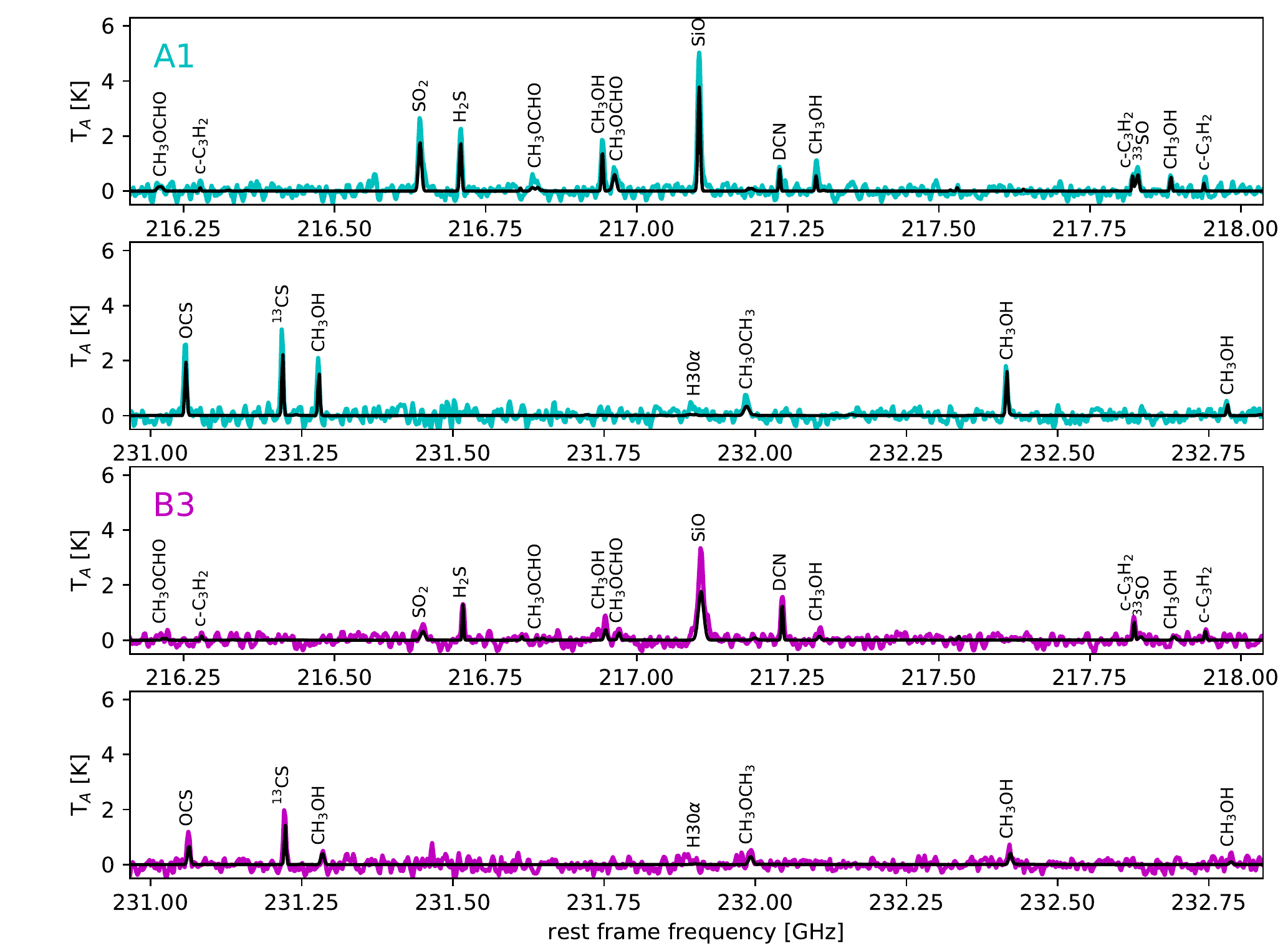}
\caption{The spectra for A1 ({\it top two panels, cyan}) and B3 ({\it bottom panels, purple}) for two Band 6 spectral windows.  The spectra were integrated over the area enclosed by the 1.3 mm continuum emission contour at the level of 50\% of the peak for the corresponding source (see Fig.~\ref{f:molimages}).  The synthetic spectra are shown in {\it black}. The molecular transitions listed in Table~\ref{t:moldata} are indicated.  \label{f:spec} }
\end{figure*}


\subsection{Spectral Line Analysis}
\label{s:specline}

The spectra for A1 and B3 are shown in Fig.~\ref{f:spec}. We detected COMs (CH$_3$OH, CH$_3$OCHO, and CH$_3$OCH$_3$), sulfur-bearing molecules (SO$_2$, H$_2$S, SiO, OCS, $^{13}$CS, and $^{33}$SO), $c$-C$_3$H$_2$, and DCN (see Table~\ref{t:moldata}).  We have carried out the line identification with the {\sc madcubaij} software (\citealt{martin2011}; \citealt{rivilla2016}), which uses the JPL\footnote{http://spec.jpl.nasa.gov/} and CDMS\footnote{http://www.astro.uni-koeln.de/cdms} molecular line databases. This software provides theoretical synthetic spectra of the different molecules under local thermodynamic equilibrium (LTE) conditions, taking into account the individual opacity of each line. To identify a molecule, the detectable lines in the observed spectra predicted by the LTE analysis must be present, with the relative intensities of the different transitions consistent with the LTE analysis.

\begin{figure}
\centering
\includegraphics[width=0.48\textwidth]{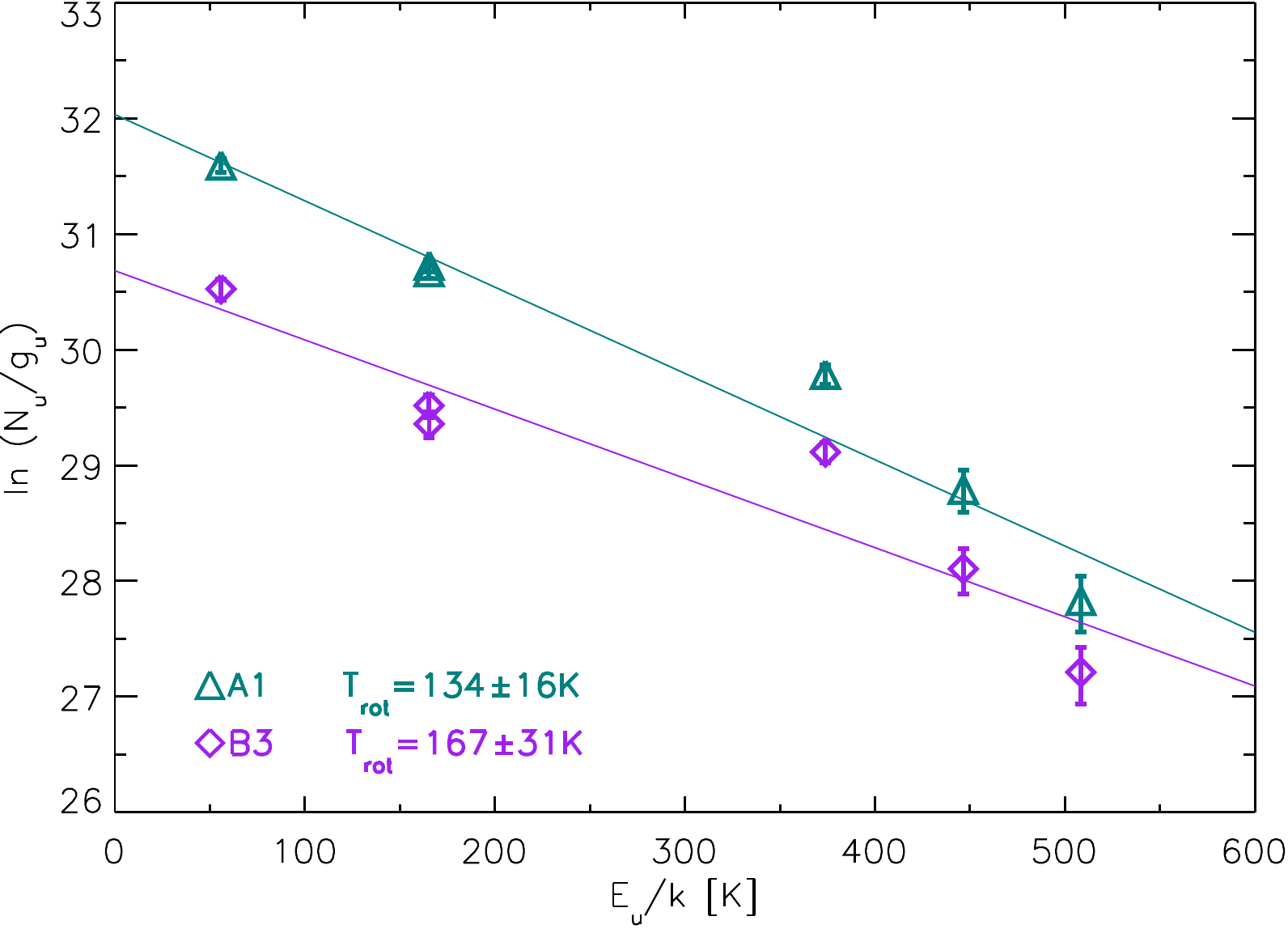}
\caption{The rotational diagram of CH$_3$OH detected toward A1 and B3.  The results of the rotational diagram analysis were used as the initial parameters for the {\sc madcubaij} modeling.   \label{f:rotdiag}}
\end{figure}

We used the results of the rotational diagram analysis (e.g., \citealt{goldsmith1999}) of CH$_3$OH as the initial parameters for the spectral line modeling using {\sc madcubaij}.  We utilized six CH$_3$OH transitions with a range of upper energy levels ($E_{\rm u}$ $\sim$56--508~K -- see Fig.~\ref{f:rotdiag}). This analysis assumes the gas is in LTE and the lines are optically thin. We obtained the rotational temperatures and total column densities ($T_{\rm rot}$, $N_{\rm rot}$) of 
($134\pm6~{\rm K}, 1.6\pm0.1\times10^{16}~{\rm cm}^{-2}$) and 
($131\pm15~{\rm K}, 6.4\pm0.8\times10^{15}~{\rm cm}^{-2}$) 
for A1 and B3, respectively.  The synthetic spectra corresponding to these parameters are shown in Fig.~\ref{f:spec}. The uncertainties for the calculated temperature and column density values are the formal fit uncertainties from the least-square fits performed by {\sc madcubaij}.  

\begin{deluxetable*}{llcccccccccc}
\centering
\tablecaption{Molecular Line Transitions Detected toward A1 and B3 \label{t:moldata}}
\tabletypesize{\scriptsize}
\tablewidth{0pt}
\tablehead{
\multicolumn{1}{c}{Species} &
\multicolumn{1}{c}{Transition} &
\colhead{Frequency} &
\colhead{$E_{\rm u}$\tablenotemark{a} } &
\multicolumn{2}{c}{Detection\tablenotemark{b}} &
\colhead{} &
\multicolumn{2}{c}{$I_{\rm int}$ (K km s$^{-1}$)\tablenotemark{c}} &
\colhead{} &
\multicolumn{2}{c}{Column Density\tablenotemark{d,e} (cm$^{-2}$)} \\
\cline{5-6}
\cline{8-9}
\cline{11-12}
\colhead{} &
\colhead{} &
\colhead{(GHz)} &
\colhead{(K)} &
\colhead{A1} & 
\colhead{B3} &
\colhead{} &
\colhead{A1} &
\colhead{B3} &
\colhead{} &
\colhead{A1} &
\colhead{B3}
}
\startdata  
CH$_{3}$OH v$_t$=0\tablenotemark{f} & 5$_{1,4}$--4$_{2,2}$ & 216.94560 & 55.87102   &       $+$ & $+$ &  & 13.8\,(0.5) & 3.4\,(0.5) & & \multirow{6}{*}{1.6\,(0.1)$\times$10$^{16}$}& \multirow{6}{*}{6.4\,(0.8)$\times$10$^{15}$}\\
CH$_{3}$OH v$_t$=1 & 6$_{1,5}$--7$_{2,6}$  & 217.299205 & 373.92517  &    $+$ & $+$  & & 8.2\,(0.2) & 2.2\,(0.2) & & &\\
CH$_{3}$OH v$_t$=0 & 20$_{1,19}$-20$_{0,20}$ & 217.88639 & 508.37582 & $+$ & $-$ & & 4.5\,(0.3) & \nodata & & &\\
CH$_{3}$OH v$_t$=0 & 10$_{2,9}$-9$_{3,6}$ & 231.28110 & 165.34719 &       $+$  &  $+$ & & 7.3\,(0.4) & 1.8\,(0.4) & & &\\
CH$_{3}$OH v$_t$=0 & 10$_{2,8}$-9$_{3,7}$ & 232.41859 & 165.40178 &       $+$ &  $+$ & & 8.1\,(0.4)& 3.7\,(0.4) & & &\\
CH$_{3}$OH v$_t$=0 & 18$_{3,16}$-17$_{4,13}$  & 232.78350 & 446.53167 &  $+$ & $+$  & & 2.5\,(0.2)& 2.7\,(0.2) & & &\\[2mm]
\hline
CH$_3$OCHO v=0 & 18$_{2,16}$--17$_{2,15}$ E & 216.83020 & 105.67781&     $+?$ &  $-$ & & 1.5\,(0.2) & \nodata &  & \multirow{6}{*}{1.1\,(0.2)$\times$10$^{15}$} & \multirow{6}{*}{$<$3.4$\times$10$^{14}$}\\
CH$_3$OCHO v=0 & 18$_{2,16}$--17$_{2,15}$ A & 216.83889 &  	105.66730  & $+?$  & $-$ & & 2.2\,(0.2) & \nodata & & &\\
CH$_3$OCHO v=0 & 20$_{1,20}$--19$_{1,19}$ E & 216.96476 & 111.49762 &     $+$ & $+?$ &  \rdelim\}{4}{0pt}[] &  \multirow{4}{*}{6.2\,(0.3)} & \multirow{4}{*}{2.4\,(0.1)} & & &\\
CH$_3$OCHO v=0 & 20$_{1,20}$--19$_{1,19}$ A & 216.96590 & 111.48055  &    $+$ & $+?$  & & && & &\\
CH$_3$OCHO v=0 & 20$_{0,20}$--19$_{0,19}$ E & 216.96625 &  	111.49826&    $+$ & $+?$ & & && & &\\
CH$_3$OCHO v=0 & 20$_{0,20}$--19$_{0,19}$ A & 216.96742 & 111.48048&       $+$ & $+?$ & & && & &\\[2mm]
\hline
CH$_3$OCH$_3$ & 13$_{0,13}$--12$_{1,12}$ EE & 231.98782& 80.92308&        $+$ & $+$& & 6.0\,(0.6) & 3.9\,(1.0) && 1.8\,(0.5)$\times$10$^{15}$ & 1.2\,(0.4)$\times$10$^{15}$\\[2mm]
\hline
c-C$_3$H$_2$ v=0 & 3$_{3,0}$--2$_{2,1} $ & 216.27876 & 19.47 &    $+$ & $+$ && 0.5\,(0.1) & 0.9\,(0.1) & & \multirow{4}{*}{3.7\,(0.7)$\times$10$^{13}$} & \multirow{4}{*}{5.6\,(1.0)$\times$10$^{13}$}\\
c-C$_3$H$_2$ v=0 & 6$_{0,6}$--5$_{1,5} $ & 217.82215 & 38.61 & $+$  & $+$ &  \rdelim\}{2}{0pt}[] & \multirow{2}{*}{2.0\,(0.3)} & \multirow{2}{*}{4.8\,(0.4)} &  & &\\
c-C$_3$H$_2$ v=0 & 6$_{1,6}$--5$_{0,5} $ & 217.82215 & 38.61 & $+$  & $+$ && & & & & \\
c-C$_3$H$_2$ v=0 & 5$_{1,4}$--4$_{2,3} $ & 217.94005 & 35.42 & $+?$ & $+$ && 1.3\,(0.2) & 2.9\,(0.3) & & &\\[2mm]
\hline
$^{33}$SO & 6$_{5}$--5$_{4}$, $F$=9/2--7/2 & 217.82718 & 34.67135 & $+$ & $+?$&  \rdelim\}{4}{0pt}[] & \multirow{4}{*}{9.2\,(0.2)} &  \multirow{4}{*}{2.9\,(0.3)}  &  & \multirow{4}{*}{1.4\,(0.2)$\times$10$^{14}$} & \multirow{4}{*}{\nodata\tablenotemark{g}}\\  
$^{33}$SO & 6$_{5}$--5$_{4}$, $F$=11/2--9/2 & 217.82983 & 34.67292 & $+$ & $+?$&& & & & & \\ 
$^{33}$SO & 6$_{5}$--5$_{4}$, $F$=13/2--11/2 & 217.83177 & 34.67502 & $+$ & $+?$ && & & & &\\ 
$^{33}$SO & 6$_{5}$--5$_{4}$, $F$=15/2--13/2 & 217.83264 & 34.67766 & $+$ &  $+?$ && & & & &\\[2mm]
\hline
SO$_2$   v=0         & 22$_{2,20}$--22$_{1,21}$ & 216.64330 & 248.44117&           $+$ & $+$ && 18.7\,(1.4) & 3.6\,(0.6) & & 3.9\,(0.4)$\times$10$^{15}$ & 9.4\,(1.7)$\times$10$^{14}$\\
H$_2$S            & 2$_{2,0}$--2$_{1,1}$     & 216.71044 & 83.98035&                      $+$ & $+$ && 11.9\,(0.8) & 7.5\,(0.3) & & 1.2\,(0.1)$\times$10$^{15}$ & 5.6\,(0.4)$\times$10$^{14}$\\
SiO                   & 5--4	                      & 217.10498 & 31.25889&                               $+$ & $+$&& 29.7\,(1.1) & 30.8\,(1.7) & & 6.9\,(0.3)$\times$10$^{13}$& 7.0\,(0.5)$\times$10$^{13}$\\
DCN v=0                &  3--2	                 & 217.238 & 20.85164 &                                $+$ & $+$ && 3.4\,(0.2) & 7.6\,(0.2) & & 1.0\,(0.1)$\times$10$^{13}$ & 1.8\,(0.1)$\times$10$^{13}$\\
OCS   v=0              & 19--18	          & 231.06098 & 110.89923&                          $+$  &  $+$ && 12.4\,(1.0) & 4.5\,(0.3) & & 8.4\,(0.9)$\times$10$^{14}$ & 4.1\,(0.3)$\times$10$^{14}$\\
$^{13}$CS        &  5--4	                 & 231.2210 &  	33.29138&                                  $+$  & $+$&& 12.8\,(0.5) & 7.9\,(0.6) & & 6.4\,(0.3)$\times$10$^{13}$& 4.1\,(0.4)$\times$10$^{13}$ \\ 
\enddata
\tablenotetext{a}{$E_u$ is the upper level energy of the transition.}
\tablenotetext{b}{The symbols in these columns indicate a detection (`$+$'), a tentative detection (`$+?$'), or a non-detection (`$-$') of a given molecular line transition.}
\tablenotetext{c}{$I_{\rm int}$ is the integrated line intensity. }
\tablenotetext{d}{The CH$_3$OH column densities and temperatures were derived based on modeling of multiple CH$_3$OH transitions ({\it see text}). Column densities of other molecules were estimated assuming the same temperature as for methanol. }
\tablenotetext{e}{For molecules with multiple transitions, all the transitions were used to estimate a total column density.}
\tablenotetext{f}{A typical line width for the strongest methanol line is $\sim$5~km~s$^{-1}$ for A1 and $\sim$8~km~s$^{-1}$ for B3.}
\tablenotetext{g}{$^{33}$SO lines for B3 are too faint to get a good fit.}
\end{deluxetable*}

Assuming that all molecular species observed toward A1/B3 are located in the same region as methanol, we estimate their column densities using the same $T_{\rm rot}$ as for methanol (see Table~\ref{t:moldata}).  To investigate how the results would change if we assumed a different value of $T_{\rm rot}$, we calculated column densities for $T_{\rm rot}$ of 100, 150, 200, and 250~K.  In general, column densities increase gradually with increasing $T_{\rm rot}$, and they are not larger than a factor of $\sim$2 than those listed in Table~\ref{t:moldata} for any molecule at $T_{\rm rot}=250$~K.  At 100~K, column densities are in general up to $\sim$20\% lower. 

A non-detection of CH$_3$OH in N\,113 by \citet{nishimura2016a}'s single dish observations can be explained by the fact that they covered Region~B only. Region~B harbors the fainter and smaller of the two sources with CH$_3$OH detection, suffering larger effects from beam dilution. The single-dish observations of \citet{wang2009} that detected methanol in N\,113 had a similar sensitivity; however, they covered both Region~A and B.   The CH$_3$OH column density estimated by \citet[$\sim$$10^{13}$ cm$^{-2}$]{wang2009} is lower than that based on the ALMA data, likely the effect of the beam dilution.  Only with ALMA can the clumpy structure of methanol be revealed. 


\section{Discussion}
\label{s:discussion}

Both A1 and B3 resemble classic hot cores -- dense condensations of molecular gas surrounding the massive star formation sites. They are compact  ($D\sim0.17$ pc) and hot  ($T_{\rm rot}\sim130~K$), as evidenced by the high excitation lines of methanol with rotation temperatures similar to Galactic hot cores at the lower end of the $T_{\rm rot}$ range (e.g., \citealt{kurtz2000}).  A1 and B3 are slightly larger than `typical' Galactic hot cores, but their number densities and column densities are consistent with those of known hot cores, they are hot enough to release ice mantles, they show emission from COMs, and are associated with masers; all these properties support the classification of A1 and B3 as hot cores. 
 
Some fainter methanol emission is detected in other locations in Region~B (see Fig.~\ref{f:molimages}); however, it is unclear whether that traces physically distinct sources or shocked lobes of the outflows. 

The detection of  SiO demonstrates that shock sputtering of dust and ice may be as important as thermal evaporation for initiating the chemical evolution in these hot cores (e.g., \citealt{charnley2000}; \citealt{viti2001}).  A SiH$_4$ (i.e., non-shock) origin for SiO is possible \citep{mackay1996}, but it requires that hot cores contain abundances of O$_2$ that are significantly above the observed upper limits obtained with {\it Herschel} (e.g., \citealt{liseau2012}; \citealt{yildiz2013}).

The detection of  CH$_3$OH  indicates that the dust was once cold enough for CO hydrogenation to proceed and the observed DCN is probably a frozen gas-phase remnant of this cold cloud chemistry \citep{brown1989}.
         
Using the CH$_3$OH and H$_2$ column densities derived above, we calculate the fractional abundance of CH$_3$OH of $(2.0\pm0.3)\times10^{-8}$ and  $(9.1\pm1.7)\times10^{-9}$ for A1 and B3, respectively.  These CH$_3$OH fractional abundances are over an order of magnitude larger than an upper limit estimated for the candidate hot core ST11 by \citet{shimonishi2016a}.  The (CH$_3$OCH$_3$, CH$_3$OCHO) fractional abundances are 
$(2.2\pm0.7,1.4\pm0.4)\times10^{-9}$ for A1 and 
$(1.7\pm0.7,<0.5)\times10^{-9}$ for B3.  
The (CH$_3$OCH$_3$, CH$_3$OCHO) abundances with respect to CH$_3$OH are 
$(0.11\pm0.03, 0.07\pm0.01)$ and 
$(0.19\pm0.07, <0.05)$ for A1 and B3, respectively. 

The abundances of COMs detected in N\,113 are comparable to those found at the lower end of the range in Galactic hot cores (e.g., \citealt{taquet2016}; \citealt{herbst2009}), when scaled by a factor of 2.5 to account for the lower metallicity in the LMC (assuming $Z_{\rm LMC}=0.4\,Z_{\odot}$).  Thus,  the chemistry of COMs detected in N\,113 is similar to that of the Galaxy, indicating that regions where they exist are likely shielded from UV radiation (see \citealt{cuadrado2017}).  Both grain reactions on warm dust and post-desorption ion--molecule chemistry could form CH$_3$OCH$_3$ and CH$_3$OCHO (\citealt{garrod2006}; \citealt{taquet2016}). 

 We have also detected several S-bearing molecules in both cores. In A1 and B3 the H$_2$S/CH$_3$OH ratios are respectively $0.075\pm0.008$ and $0.088\pm0.013$, whereas the OCS/CH$_3$OH ratios are $0.053\pm0.007$ and $0.064\pm0.009$.  These ratios are almost indistinguishable between A1 and B3, supporting the idea that these molecules were formed in ices during the N113 prestellar phase.  OCS is known to be present  in interstellar ices \citep{boogert2015} and, although H$_2$S has not yet been confirmed,  cold  grain-surface chemistry,  involving reactions of S atoms with atomic hydrogen and CO (or oxidation of CS),   is likely to be the origin of these molecules in the N113 cores. 

SO and SO$_2$ could also originate from ice chemistry (they are detected in cometary ices; \citealt{calmonte2016}), but they may also form in post-shock gas that is enriched in OH through neutral--neutral reactions \citep{hartquist1980}.  Alternatively, chemical models show that all the S-bearing molecules, including SO and SO$_2$,   could be  formed in post-evaporation gas-phase reactions starting  from evaporated ices containing  H$_2$S \citep{charnley1997}.  In this case,  comparison of calculated and observed  SO/SO$_2$, H$_2$S/SO$_2$ and H$_2$S/SO ratios could be used as a chemical clock for the age of the core. We will explore this issue in a future publication.

 Finally,  $c$-C$_3$H$_2$ could either be formed by ion--molecule chemistry in the cool ambient medium or, like CCH, be in a PDR region, where it may be produced in the photodestruction of larger refractory organic (PAH) molecules \citep{guzman2014}.

\section{Conclusions}

We see the effective formation of COMs under sub-solar metallicity conditions in two hot cores discovered in the LMC.  The COMs observed in the N\,113 hot cores could either originate from grain surface chemistry or in post-desorption gas chemistry.  Of the S-bearing molecules detected, it is likely that  H$_2$S and OCS formed on grains; SO and SO$_2$  also could have originated from ice chemistry or formed in gaseous neutral reactions.  The presence of SiO indicates that shock chemistry may also be playing a role. Future observations of LMC cores and detailed chemical modeling will be necessary to determine the relative contribution from each process.  Studying the chemistry of the interstellar medium as a function of metallicity is important to understand the chemical evolution of the
Universe. 

\acknowledgements 

We thank the anonymous referee for insightful comments and suggestions which helped us improve the paper.  The work of M.S. was supported by an appointment to the NASA Postdoctoral Program at the Goddard Space Flight Center, administered by Universities Space Research Association under contract with NASA.  S.C. acknowledges the support from the NASA's Emerging Worlds Program.  S.Z. was supported by NAOJ ALMA Scientific Research Grant Number 2016-03B.  J.L.W. acknowledges support from the German Research Council (SFB 881 ``The Milky Way System'', subproject P1). M.M. was supported by NSF grant 1312902.  The National Radio Astronomy Observatory is a facility of the National Science Foundation operated under cooperative agreement by Associated Universities, Inc. This paper makes use of the following ALMA data: ADS/JAO.ALMA\#2015.1.01388.S.  ALMA is a partnership of ESO (representing its member states), NSF (USA) and NINS (Japan), together with NRC (Canada), NSC and ASIAA (Taiwan), and KASI (Republic of Korea), in cooperation with the Republic of Chile. The Joint ALMA Observatory is operated by ESO, AUI/NRAO and NAOJ.



\begin{thebibliography}{}
\bibitem[Abdo et al.(2010)]{abdo2010} Abdo, A.~A., Ackermann, M., Ajello, M., et al.\ 2010, \aap, 512, A7 
\bibitem[Acharyya \& Herbst(2015)]{acharyya2015} Acharyya, K., \& Herbst, E.\ 2015, \apj, 812, 142
\bibitem[Boogert et al.(2015)]{boogert2015} Boogert, A.~C. Adwin, Gerakines, P.~A., Whittet, D.~C.~B.\ 2015, \araa, 53, 541
\bibitem[Brown et al.(1988)]{brown1988} Brown, P.~D., Charnley, S.~B., \& Millar, T.~J.\ 1988, \mnras, 231, 409 
\bibitem[Brown \& Millar(1989)]{brown1989} Brown, P.~D., \& Millar, T.~J.\ 1989, \mnras, 237, 661
\bibitem[Calmonte et al.(2016)]{calmonte2016} Calmonte, U., Altwegg, K., Balsiger, H., et al.\ 2016, \mnras, 462, 253
\bibitem[Carlson et al.(2012)]{carlson2012} Carlson, L.~R., Sewi{\l}o, M., Meixner, M., Romita, K. A., \& Lawton, B.\ 2012, \aap, 542, A66
\bibitem[Charnley et al.(1992)]{charnley1992} Charnley, S.~B., Tielens, A.~G.~G.~M., Millar, T.~J.\ 1992, \apj, 399, 71
\bibitem[Charnley(1997)]{charnley1997} Charnley, S.~B. 1997, \apj, 481, 396
\bibitem[Charnley \& Kaufman(2000)]{charnley2000} Charnley, S.~B., \& Kaufman, M. J.\ 2000, \apj, 529, 111
\bibitem[Charnley \& Rogers(2008)]{charnley2008} Charnley, S.~B., \& Rodgers, S.~D.\ 2008, \ssr, 138, 59
\bibitem[Cuadrado et al.(2017)]{cuadrado2017} Cuadrado, S., Goicoechea, J. R., Cernicharo, J., et al.\ 2017,A\&A, 603, 124
\bibitem[Ehrenfreund \& Charnley(2000)]{ehrenfreund2000} Ehrenfreund, P., \& Charnley, S.~B.\ 2000, \araa, 38, 427
\bibitem[Ellingsen et al.(2010)]{ellingsen2010} Ellingsen, S.~P., Breen, S.~L., Caswell, J.~L., Quinn, L.~J., \& Fuller, G. A.\ 2010, \mnras, 404, 779
\bibitem[Fontani et al.(2011)]{fontani2011} Fontani, F., Palau, A., Caselli, P., et al.\ 2011, \aap, 529, 7 
\bibitem[Galliano et al.(2011)]{galliano2011} Galliano, F., Hony, S., Bernard, J.-P., et al.\ 2011, \aap, 536, A88
\bibitem[Garrod et al.(2006)]{garrod2006} Garrod, R.~T., \& Herbst, E.\ 2006, \aap, 457, 927
\bibitem[Goldsmith \& Langer(1999)]{goldsmith1999} Goldsmith, P.~F., \& Langer, W.~D.\ 1999, \apj, 517, 209
\bibitem[Green et al.(2008)]{green2008} Green, J.~A., Caswell, J.~L., Fuller, G.~A., et al.\ 2008, \mnras, 385, 948
\bibitem[Gruendl \& Chu(2009)]{gruendl2009} Gruendl, R.~A., \& Chu, Y.\ 2009, \apjs, 184, 172
\bibitem[Guzm{\'a}n et al.(2014)]{guzman2014} Guzm{\'a}n, V.~V., Pety, J., Gratier, P., et al. 2014, Faraday Discussions, 168, 103
\bibitem[Hartquist et al.(1980)]{hartquist1980} Hartquist, T.~W., Dalgarno, A., Oppenheimer, M.\ 1980, \apj, 236, 182
\bibitem[Heikkil{\"a} et al.(1999)]{heikkila1999} Heikkil{\"a}, A.; Johansson, L.~E.~B.; Olofsson, H.\ 1999, \aap, 344, 817
\bibitem[Henize(1956)]{henize1956} Henize, K.~G.\ 1956, \apjs, 2, 315
\bibitem[Herbst \& van Dishoeck(2009)]{herbst2009} Herbst, E. \& van Dishoeck. E.F.\ 2009, \araa,  47, 427
\bibitem[Hollenbach \& Tielens (1999)]{hollenbach1999} Hollenbach, D.~J. \& Tielens, A.~G.~G.~M.\ 1999, Reviews of Modern Physics, 71, 173
\bibitem[Kurtz et al.(2000)]{kurtz2000} Kurtz, S., Cesaroni, R., Churchwell, E., Hofner, P., \& Walmsley, C. M.,\ 2000, Protostars and Planets IV (Book - Tucson: University of Arizona Press; eds Mannings, V., Boss, A.P., Russell, S. S.), p. 299-326
\bibitem[Liseau et al.(2012)]{liseau2012}	Liseau, R., Goldsmith, P. ~F.; Larsson, B., Pagani, L., et al.\ 2012, \aap, 541, 73
\bibitem[Mackay(1996)]{mackay1996} Mackay, D.~D.~S.\ 1996, \mnras, 278, 62
\bibitem[Mart{\'{\i}}n et al.(2011)]{martin2011} Mart{\'{\i}}n, S., Krips, M., Mart{\'{\i}}n-Pintado, J., et al.\ 2011, \aap, 527, A36
\bibitem[McMullin et al.(2007)]{mcmullin2007} McMullin, J.~P., Waters, B., Schiebel, D., Young, W., \& Golap, K.\ 2007, in Astronomical Society of the Pacific Conference Series, Vol. 376, Astronomical Data Analysis Software and Systems XVI, ed. R. A. Shaw, F. Hill, \& D. J. Bell, 127
\bibitem[Meixner et al.(2006)]{meixner2006} Meixner, M., Gordon, K.~D., Indebetouw, R., et al.\ 2006, \aj, 132, 2268
\bibitem[Nishimura et al.(2016a)]{nishimura2016a}  Nishimura, Y., Shimonishi, T.,  Watanabe, Y., et al.\ 2016, \apj, 818, 161 
\bibitem[Nishimura et al.(2016b)]{nishimura2016b}  Nishimura, Y., Shimonishi, T.,  Watanabe, Y., et al.\ 2016, \apj, 829, 94 
\bibitem[Oliveira et al.(2006)]{oliveira2006} Oliveira, J.~M., van Loon, J.~T., Stanimirov{\'i}c, S., \& Zijlstra, A.~A.\ 2006, \mnras, 372, 1509
\bibitem[Oliveira et al.(2011)]{oliveira2011} Oliveira, J.~M., van Loon, J.~Th., Sloan, G.~C., et al. 2011\, \mnras, 411, 36
\bibitem[Pietrzy{\'n}ski et al.(2013)]{pietrzynski2013} Pietrzy{\'n}ski, G., Graczyk, D., Gieren, W., et al.\ 2013, \nat, 495, 76
\bibitem[Rivilla et al.(2016)]{rivilla2016} Rivilla, V.~M., Fontani, F., Beltr{\'a}n, M.~T., et al.\ 2016, \apj, 826, 161
\bibitem[Seale et al.(2012)]{seale2012} Seale, J.~P., Looney, L.~W., Wong, T., et al.\ 2012, \apj, 751, 42
\bibitem[Sewi{\l}o et al.(2010)]{sewilo2010} Sewi{\l}o, M., Indebetouw, R., Carlson, L.~R., et al.\ 2010, \aap, 518, L73
\bibitem[Shimonishi et al.(2016a)]{shimonishi2016a} Shimonishi, T., Onaka, T., Kawamura, A., \& Aikawa, Y.\ 2016, \apj, 827, 72 
\bibitem[Shimonishi et al.(2016b)]{shimonishi2016b} Shimonishi, T., Dartois, E., Onaka, T., \& Boulanger, F.\ 2016, \aap, 585, A107 
\bibitem[Sinclair et al.(1992)]{sinclair1992} Sinclair, M.~W., Carrad, G.~J., Caswell, J.~L., Norris, R.~P., \& Whiteoak, J.~B.\ 1992, \mnras, 256, 33
\bibitem[Smith \& MCELS Team(1998)]{smith1998} Smith, R. C., \& MCELS Team.\ 1998, \pasa, 15, 163
\bibitem[Taquet et al.(2016)]{taquet2016} Taquet, V.,  Wirstr{\"o}m, E.~S., \& Charnley, S.~B.\ 2016, \apj, 821, 46
\bibitem[van Loon et al.(2010)]{vanloon2010}	 van Loon, J.~Th., Oliveira, J.~M., Gordon, K.~D., Sloan, G.~C., Engelbracht, C.~W.\ 2010, \aj, 139, 1553
\bibitem[Viti et al.(2001)]{viti2001} Viti, S., Caselli, P., Hartquist, T.~W., \& Williams, D.~A.\ 2001, \aap, 370, 1017
\bibitem[Wang et al.(2009)]{wang2009} Wang, M., Chin, Y.-N., Henkel, C., Whiteoak, J.~B., \& Cunningham, M. 2009, \apj, 690, 580
\bibitem[Ward et al. (2016)]{ward2016} Ward, J.~L., Oliveira, J.~M., van Loon, J.~T., \& Sewi{\l}o, M.\ 2016, \mnras, 455, 2345
\bibitem[Watanabe \& Kouchi(2008)]{watanabe2008} Watanabe, N., \& Kouchi, A. 2008, Progress in Surface Science\ 2008, vol. 83, 10, 439
\bibitem[Westerlund(1997)]{westerlund1997} Westerlund, B. E.\, 1997, The Magellanic Clouds, (New York: Cambridge Univ.
Press
\bibitem[Whiteoak \& Gardner(1986)]{whiteoak1986} Whiteoak, J.~B., \& Gardner, F.~F.\ 1986, \mnras, 222, 513
\bibitem[Wong et al.(2011)]{wong2011} Wong, T., Hughes, A., Ott, J., et al.\ 2011, \apjs, 197, 16
\bibitem[{Y{\i}ld{\i}z} et al.(2013)]{yildiz2013} {Y{\i}ld{\i}z}, U.~A., Acharyya, K., Goldsmith, P.~F., et al.\ 2013, \aap, 558, 58
\end{thebibliography}
\end{document}